\documentclass[letterpaper]{article}

\usepackage[T1]{fontenc}
\usepackage[utf8]{inputenc}
\usepackage{lmodern}
\usepackage[english]{babel}
\usepackage[
 top=2.5cm,bottom=3.2cm,left=3cm,right=3cm,
  marginparwidth=1.75cm,footskip=15mm
]{geometry}

\usepackage{microtype}
\usepackage{amsmath,amssymb,amsfonts}
\usepackage{graphicx}
\usepackage{authblk}
\usepackage{float}     
\usepackage{placeins}  
\usepackage{hanging}   

\usepackage[colorlinks=true, allcolors=blue]{hyperref}

\title{Interval-Censored Survival Analysis of Grapevine Phenology: Thermal Controls on Flowering and Fruit Ripening}

\author[1]{Sara Behnamian}
\author[2]{Fatemeh Fogh}

\affil[1]{Globe Institute, University of Copenhagen, \mbox{Øster Voldgade 5--7}, 1350 Copenhagen K, Denmark\\
\texttt{sara.behnamian@sund.ku.dk}}
\affil[2]{Department of Mathematical Sciences, Florida Atlantic University, Boca Raton, FL 33431, USA\\
\texttt{ffogh2021@fau.edu}}

\date{} 

\begin{document}
\maketitle
\begin{abstract}
European grapevine (\textit{Vitis vinifera} L.) is a climate–sensitive perennial whose flowering and ripening govern yield and quality. Phenological records from monitoring programs are typically collected at irregular intervals, so true transition dates are interval-censored, and many site-years are right-censored. We develop a reproducible workflow that treats phenology as a time–to–event outcome: Status \& Intensity observations from the USA–NPN are converted to interval bounds, linked to NASA POWER daily weather, and analyzed with parametric accelerated failure time (AFT) models (Weibull and log–logistic). To avoid outcome–dependent bias from aggregating weather up to the event date, antecedent conditions are summarized in fixed pre–season windows and standardized; quality–control filters ensure adequate within–window data coverage.

Applied to flowering and ripening of \textit{V.~vinifera}, the framework yields interpretable time–ratio effects and publication-ready tables and figures. Warmer pre–season conditions are associated with earlier ripening, whereas flowering responses are modest and uncertain in these data; precipitation plays, at most, a secondary role. The approach demonstrates how interval-censored survival models with exogenous weather windows can extract robust climate signals from citizen-science phenology while preserving observation uncertainty, and it generalizes readily to other species and networks.
\end{abstract}

\section{Introduction}
European grapevine (\textit{Vitis vinifera} L.) is among the world’s most valuable perennial crops, cultivated on more than seven million hectares and producing over 77~Mt annually across six continents (OIV, 2023). Beyond its economic importance, grapevine is an ideal model for climate–phenology studies: its phenophases are conspicuous and agronomically decisive, long historical records exist for many regions and cultivars, and modern monitoring networks continue to amass large, standardized datasets (Wolkovich et al., 2012; Parker et al., 2020). Anticipating how flowering and ripening shift under ongoing warming is central to adaptation because phenological timing governs frost exposure, sink–source balance, berry composition, and ultimately which cultivar–region combinations remain viable (Hannah et al., 2013; van Leeuwen \& Darriet, 2016; Morales‐Castilla et al., 2020).

Temperature is the dominant driver of grapevine development from dormancy release through harvest, with forcing usually summarized as growing degree days (GDD) accumulated above a base temperature near 10~\(^\circ\)C for \textit{V.~vinifera} (Bonhomme, 2000; McMaster \& Wilhelm, 1997; Moncur et al., 1989; Williams et al., 1985). Thermal‐time models capture much of the variance in the timing of budburst, flowering, and véraison across sites and cultivars, yet their extrapolation in novel climates is uncertain and can be confounded by heat extremes, water status, and cultivar‐specific base temperatures (Parker et al., 2011; García de Cortázar‐Atauri et al., 2009; Zapata et al., 2017). These secondary factors—particularly water balance—modulate rates during ripening and can either accentuate or buffer temperature effects depending on stress severity and timing, reinforcing the need for analyses that quantify multiple antecedent drivers (van Leeuwen \& Darriet, 2016; Parker et al., 2020).

Methodologically, phenology presents nontrivial statistical challenges. Field observers visit plants intermittently; thus the true transition time is known only to lie between the last “no” and first “yes” observations—an interval‐censored outcome (Sun, 2006; Turnbull, 1976). Using the first observed date as the response produces frequency‐dependent bias and discards years with no detection (right censoring), both of which can distort climate sensitivities (Moussus et al., 2010; Clark et al., 2014; Polgar \& Primack, 2011). Survival analysis addresses these issues directly by modeling time to event with censoring, and accelerated failure time (AFT) models are especially interpretable for phenology because covariate effects are expressed as \emph{time ratios}: values below 1 indicate acceleration (earlier timing) and above 1 indicate delay (Wei, 1992; Hosmer et al., 2008; Klein \& Moeschberger, 2003; Collett, 2015). Recent applications show that accounting for censoring strengthens and clarifies climate signals compared with regressions on first dates (Roberts, 2008; Calinger et al., 2013; Jarrad et al., 2018).

Despite recognition of these statistical issues, there remains no comprehensive framework for applying interval-censored survival analysis to phenological data. Researchers face several persistent barriers: (i) limited guidance on converting standard monitoring records into interval- and right-censored formats; (ii) lack of clear workflows for constructing covariates in ways that avoid outcome-dependent endogeneity; (iii) few demonstrations that compare biased versus unbiased approaches using real observational data; and (iv) absence of accessible tools that integrate these steps into a reproducible pipeline. As a result, many studies continue to rely on first-observed dates with standard regression models, potentially biasing inference on climate sensitivities and limiting the interpretability of results.

A further pitfall is endogeneity in weather metrics constructed “to the event.” Cumulative heat or precipitation integrated up to the observed date is mechanically correlated with the outcome, inflating associations and complicating causal interpretation—especially when the date itself is interval‐censored (García de Cortázar‐Atauri et al., 2009). To mitigate this, exogenous fixed windows that precede typical phenology (e.g., DOY~1–120 before flowering) can be used to summarize antecedent conditions while avoiding outcome dependence, at the cost of requiring careful window selection and collinearity checks (Čufar et al., 2011; Legave et al., 2013).

Here we leverage standardized presence/absence observations from the USA National Phenology Network (USA–NPN) for cultivated grapevines and pair them with daily meteorology from NASA POWER. USA–NPN’s Status \& Intensity protocol generates exactly the information needed for interval‐censored modeling—explicit “no/yes” sequences on marked individuals with metadata suitable for quality control (Denny et al., 2014; Rosemartin et al., 2018; USA–NPN, 2024). Our objectives are to: (i) develop and validate a generalizable framework for interval-censored survival analysis of phenological data; (ii) diagnose and quantify endogeneity bias arising from conventional “to-event” covariate construction; (iii) provide an open, reproducible pipeline (data ingestion, interval construction, QC, exogenous covariate design, model fitting, and reporting) that yields publication-ready outputs; and (iv) illustrate the framework with grapevine flowering and ripening as an applied case study, while emphasizing generalization to other phenological systems and networks.

From an applied perspective, earlier ripening under warmer pre–seasons has direct consequences for harvest logistics, berry composition, and wine style, with potential shifts in sugar–acid balance, phenolics, and aromatic precursors (van Leeuwen \& Darriet, 2016; Parker et al., 2020). Because cultivars differ in thermal requirements and base temperatures, climate warming can decouple current cultivar–region pairings from their historically optimal phenological windows, altering frost risk, heat exposure, and water demand (Hannah et al., 2013; Morales–Castilla et al., 2020). By delivering time–ratio estimates that link antecedent heat and moisture to the timing of flowering and ripening while properly accounting for censoring and avoiding outcome–dependent bias, our framework provides a decision‐oriented basis for adaptation—e.g., cultivar choice, pruning/irrigation scheduling, canopy management, or, where feasible, site shifts within regions. In this way, the methodology complements thermal‐time models by quantifying how exogenous pre–season conditions advance or delay key phenophases in a manner directly interpretable for viticultural planning.
This work therefore goes beyond previous phenology–climate studies by uniting 
interval‐censored survival models with a transparent, scriptable pipeline. 
Where earlier analyses often focused on specific species, sites, or 
statistical demonstrations, our contribution is to integrate data processing, 
covariate design, model fitting, and reproducible output into one 
generalizable framework. In doing so, we provide not only methodological rigor 
but also practical tools for researchers and practitioners working across 
phenological networks and taxa.

In sum, our study contributes a practical, end-to-end workflow for interval-censored phenology analysis. The framework links raw observational data to survival models through transparent steps that avoid common statistical pitfalls, and it produces effect estimates directly interpretable for ecological and agricultural applications. While demonstrated here for grapevine, the same pipeline applies broadly to phenological datasets across taxa and monitoring networks.

\begin{figure}[!htbp]
  \centering
  \includegraphics[width=0.7\linewidth]{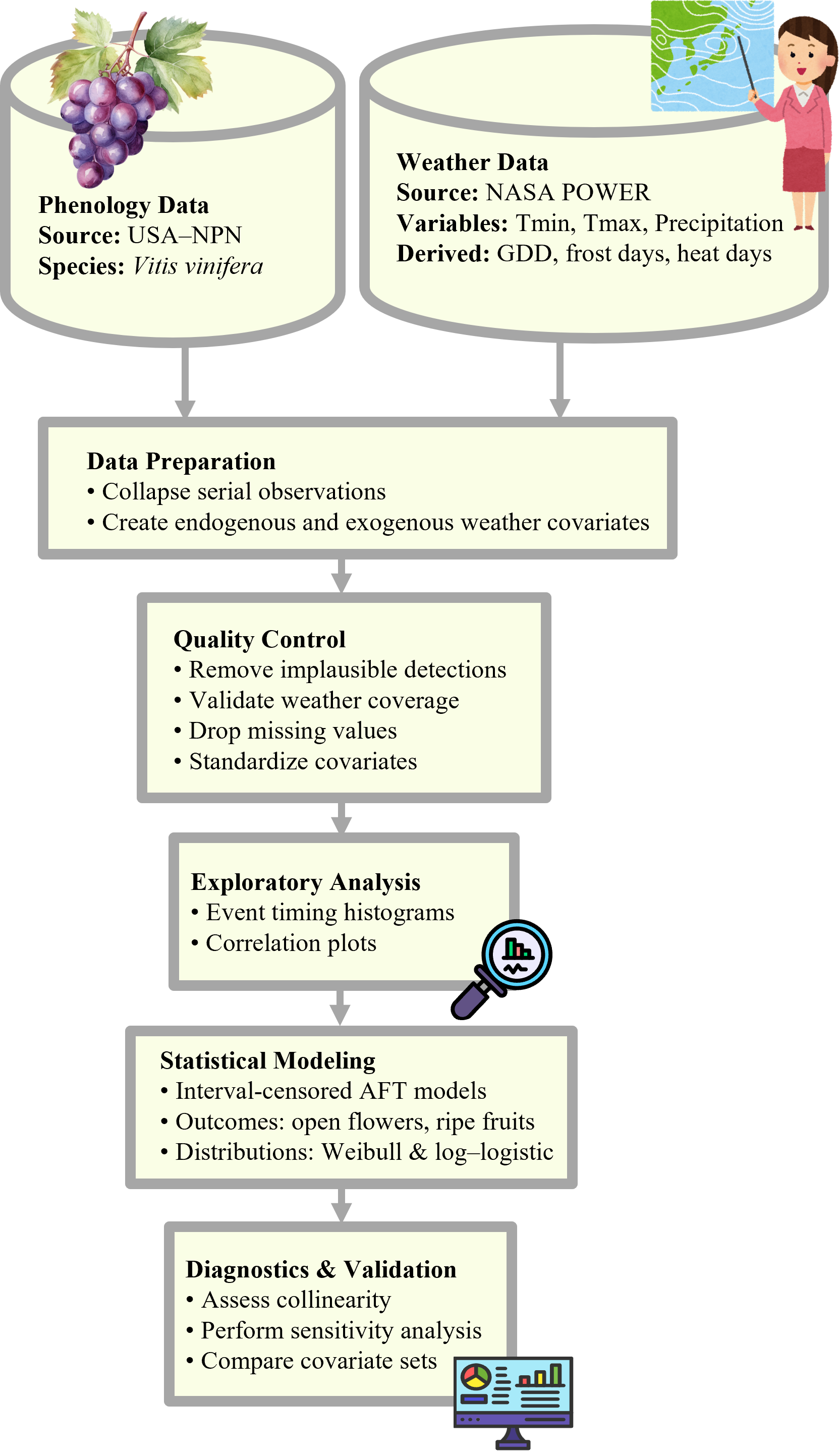}
  \caption{Analysis workflow: data sources, preparation, QC, exploratory analysis, interval–censored AFT modeling, and diagnostics/validation.}
  \label{fig:workflow}
\end{figure}

As an overview, our end-to-end workflow is shown in Figure~\ref{fig:workflow}.

\FloatBarrier

\section{Methods}

\section*{Data}

\subsection*{Study Species and Phenological Observations}
We monitored 7 individual European grapevine (\textit{Vitis vinifera} L.) plants across 5 sites, yielding 41{,}274 phenophase status observations between 30~March~2012 and 16~May~2025. These observations were collected by 16 observers using the USA–NPN Status \& Intensity protocol (Denny et al., 2014; Rosemartin et al., 2018; USA–NPN, 2024).

Although 41{,}274 daily records were available, interval‐censoring requires collapsing them to one survival unit per plant–site–year. This aggregation yields 50 usable analytical records across the study period (not all plants were monitored in all years), which form the basis of subsequent survival analyses.

\subsection*{Phenophase Definitions and Sample Size}
Our survival endpoints are \emph{Open flowers} (Phenophase Definition ID~501) and \emph{Ripe fruits} (ID~390), defined in USA–NPN documentation (``one or more open, fresh flowers are visible''; ``one or more fruits are fully ripe or have passed the point of ripeness'') (Rosemartin et al., 2018; USA–NPN, 2024). We analyzed 4{,}140 records for \emph{Open flowers} (present = 1{,}247; absent = 2{,}893) and 4{,}134 for \emph{Ripe fruits} (present = 1{,}089; absent = 3{,}045); 29 ``uncertain'' records were excluded. Intensity (percent‐binned) is reported for 22.0\% of all observations overall, 3.3\% for \emph{Open flowers}, and 20.5\% for \emph{Ripe fruits} (otherwise a sentinel value denotes not reported) (USA–NPN, 2024). For reproducibility and unambiguous joins to USA–NPN dictionaries, endpoints are specified by their numeric identifiers (501, 390).

\subsection*{Event Definition and Censoring}
We derived time‐to‐event outcomes for the phenophases \emph{Open flowers} and \emph{Ripe fruits} from USA–NPN Status \& Intensity observations (USA–NPN, 2024; Rosemartin et al., 2018). Calendar dates were converted to year and day‐of‐year (DOY). Phenophase status was coded as present (=1) or absent (=0), and observations marked uncertain (=$-1$) were excluded. Within each plant–site–year, records were ordered by date to define an interval \(T \in (L, R]\) for the first occurrence of the target phenophase: \(R\) is the DOY of the first present observation and \(L\) is the DOY of the most recent absent observation preceding \(R\). If the first observation of a year was present, \(L\) was set to one day before the first observed DOY. Years with no present observation were treated as right‐censored at the last observed DOY (event = 0). Endpoints were referenced by their USA–NPN numeric identifiers to ensure unambiguous linkage to official definitions (Open flowers: 501; Ripe fruits: 390). The procedure yields one record per plant–site–year containing \(L\), \(R\), an event indicator, the number of visits contributing to the interval, and the first and last observed DOY; intensity responses were not used to define events (Turnbull, 1976; Sun, 2006; Klein \& Moeschberger, 2003).

\subsection*{Weather Covariates}
Daily meteorological data were obtained from NASA’s Prediction Of Worldwide Energy Resources (POWER) project for each site across its observed phenology window (NASA POWER Project, 2025). Point queries at site latitude and longitude returned minimum and maximum 2-m air temperature and bias‐corrected daily precipitation (variables \texttt{T2M\_MIN}, \texttt{T2M\_MAX}, \texttt{PRECTOTCORR}) in $^\circ$C and mm~day$^{-1}$. Requests were issued in calendar‐year segments, merged across parameters on the union of available dates, deduplicated by date, and trimmed to the site‐specific observation window. Weather coverage exactly matched each site’s observed phenology span, as determined from site–year summaries. Coordinates were validated prior to querying. No temporal gap filling or spatial interpolation was performed; missing dates remain missing.

From temperature we derived daily growing degree days with a 10~$^\circ$C base as
\[
\mathrm{GDD}_{10,d}=\max\!\left(\frac{T_{\max,d}+T_{\min,d}}{2}-10,\ 0\right).
\]
The meteorological series were then aligned to the phenology records by site and calendar date to enable joint analyses of event timing with temperature and precipitation drivers (Bonhomme, 2000; McMaster \& Wilhelm, 1997).

\subsection*{Covariate Aggregation and Linkage}
Daily weather was aggregated to a phenology‐relevant cutoff for each plant–site–year. The cutoff DOY was defined as the first detection of the endpoint ($R$, the right bound of the interval) when an event occurred; otherwise it was the last observed DOY in that year. For each site–year, NASA POWER daily series were ordered by date and cumulative summaries were taken up to and including the cutoff.

From these series we derived antecedent covariates: cumulative GDD\(_{10}\) to cutoff; cumulative precipitation to cutoff; mean $T_{\min}$ and mean $T_{\max}$ to cutoff; counts of frost days ($T_{\min}\le 0^\circ$C) and heat days ($T_{\max}\ge 35^\circ$C) to cutoff; and the number of weather days contributing together with the maximum DOY used. Joins used site and calendar‐year keys, and no temporal gap filling was applied; missing weather dates remain missing.

\subsection*{Exploratory Analyses and Quality Control}
We conducted exploratory summaries on the merged phenology–weather dataset to characterize event occurrence and timing and to screen for biologically implausible records. For each endpoint (\emph{Open flowers}; \emph{Ripe fruits}), we tallied plant–site–year units, estimated the proportion with an observed event, and computed the median DOY of first detection ($R$) among event years. To assess thermal forcing at first detection, we calculated Pearson correlations between $R$ and cumulative GDD\(_{10}\) accumulated to the endpoint‐specific cutoff; correlations were evaluated on event years with pairwise deletion for missing covariates. As a plausibility screen, ripe‐fruit detections with $R<120$~DOY were flagged as likely carry‐over from the previous season at the study latitudes; summaries were reported before and after excluding these cases. We additionally summarized site‐level event rates. Distributions of $R$ (events only) were displayed as histograms with medians, and the relationship between cumulative heat and timing was visualized with scatterplots of GDD versus $R$, overlaying an ordinary least‐squares trend line when $|r|>0.2$. Weather histories were used exactly as aggregated to the cutoff with no temporal gap filling. Across the five study sites, daily NASA POWER coverage spanned 2012–2025 with continuous records over each site’s observation window.

\subsection*{Interval-Censored Accelerated Failure Time Models}
We modeled the timing of first occurrence (\(T\), DOY) for \textit{Open flowers} and \textit{Ripe fruits} using parametric AFT models with interval censoring. For each plant–site–year, the event time entered the likelihood as an interval \((L, R]\); years without a detection were treated as right‐censored by setting \(R=\infty\). To ensure valid bounds we clipped \(L\) to \([1,366]\) and, where a finite \(R\le L\) occurred, replaced \(R\) with \(L+\varepsilon\) (\(\varepsilon=10^{-6}\)). Based on biological screening, putative carry‐over fruit detections were excluded for ripening ($R<120$~DOY). Predictor variables summarized antecedent conditions to the phenology‐specific cutoff and included cumulative GDD\(_{10}\), cumulative precipitation, and counts of frost days ($T_{\min}\le 0^\circ$C) and heat days ($T_{\max}\ge 35^\circ$C); all covariates were standardized (mean~0, SD~1) prior to fitting. We fit Weibull and log–logistic AFT specifications by maximum likelihood, reporting coefficients on the log‐time scale and interpreting effects as time ratios (\(\mathrm{TR}=\exp\beta\)). We additionally obtained model‐based medians at mean covariate values. Unless stated otherwise, all four antecedent covariates aggregated to the cutoff entered jointly after $z$‐scoring. Rows with incomplete interval bounds or any missing antecedent covariate were excluded listwise (Wei, 1992; Hosmer et al., 2008; Collett, 2015).

\subsection*{Validation and Sensitivity Analysis}
We validated the merged phenology–weather dataset by requiring complete meteorological coverage through the phenology cutoff used to aggregate covariates (i.e., the first detection day \(R\) for event years or the last observation day for censored years). For \emph{Ripe fruits}, implausibly early detections consistent with prior–season carry–over were excluded ($R<120$~DOY). Interval bounds were rechecked ($L\in[1,366]$; finite $R>L$ or $R=\infty$ for right–censoring), antecedent covariates were standardized, and Weibull and log–logistic AFT models were refit on the validated set. As a sanity check, we computed the Pearson correlation between $R$ and cumulative GDD\(_{10}\) to the cutoff among event years only. Model summaries report time ratios (TR $=\exp\beta$) with 95\% confidence intervals and model‐based medians at mean covariates.

\subsection*{Covariate Diagnostics and Exogenous Weather}
We first screened the four antecedent covariates previously aggregated to the phenology cutoff—cumulative GDD, cumulative precipitation, frost‐day count, and heat‐day count—for collinearity and stability. Years without a detection were treated as right‐censored ($R=\infty$), and implausibly early ripening detections ($R<120$~DOY) were excluded.

To avoid potential endogeneity introduced by aggregating weather up to an outcome‐dependent cutoff, we derived \emph{exogenous} pre–season features that end before typical detections: DOY~1–120 for flowering and DOY~1–180 for ripening. For each site–year we summarized daily meteorology to obtain $\mathrm{GDD}_{\text{pre}}$, $\mathrm{PRCP}_{\text{pre}}$, $\overline{T}_{\min,\text{pre}}$, $\overline{T}_{\max,\text{pre}}$, $\mathrm{frost}_{\text{pre}}$ (days with $T_{\min}\le 0^\circ$C), and $\mathrm{heat}_{\text{pre}}$ (days with $T_{\max}\ge 35^\circ$C); primary totals were standardized to $z$–scores (e.g., $\mathrm{GDD}_{\text{pre},z}$). We then fit univariate interval–censored AFT models (Weibull and log–logistic) using $\mathrm{GDD}_{\text{pre},z}$ as the sole predictor and reported $\mathrm{TR}=\exp(\beta)$ with 95\% Wald CIs. For illustration of collinearity effects, we also fit an optional joint Weibull AFT with the current to–cutoff covariates after $z$–scoring (Čufar et al., 2011; Legave et al., 2013).

\subsection*{Bivariate Pre–Season Models with Confidence Intervals}
Daily weather was aggregated into fixed pre–season windows that end well before typical detection: DOY~1–120 for \emph{Open flowers} and DOY~1–180 for \emph{Ripe fruits}. For each site–year we summed GDD\(_{10}\) and precipitation within the window and computed coverage as observed/expected days; records with coverage $<0.70$ were excluded. These window features were merged to the interval‐censored survival tables by site and year. Survival bounds were sanitized by clipping $L$ and finite $R$ to $[1,366]$, setting $R=\infty$ when missing (right‐censoring), and replacing degenerate cases with $R=L+10^{-6}$. Pre–season GDD and precipitation were standardized to $z$‐scores within the analysis set.

For each endpoint we fit bivariate AFT models with Weibull and log–logistic distributions using interval censoring. Covariates were the two pre–season $z$‐scored predictors: $\mathrm{GDD}_{\text{pre}}$ and $\mathrm{precip}_{\text{pre}}$. Effects are reported on the time–ratio (TR) scale, defined as the multiplicative change in the median event time per $+1$~SD increase in a covariate (TR$<1$ earlier, TR$>1$ later). Ninety‐five percent confidence intervals were computed as $\exp(\hat\beta \pm 1.96\,\mathrm{SE})$.

\section{Results}
\subsection*{Interval construction and detection rates}
From the USA–NPN observations we constructed \(n=50\) plant–site–year records across the study period. Of these, 30 had observed events for \emph{Open flowers} and 37 for \emph{Ripe fruits}, corresponding to event rates of 0.60 and 0.76, respectively. Median timing among event years was DOY~150 for \emph{Open flowers} and DOY~205 for \emph{Ripe fruits} (Figure~\ref{fig:timing}).

\subsection*{Weather coverage and antecedent conditions}
NASA POWER retrieval yielded 8{,}354 site–days of weather across five sites (2012–03–30 to 2025–05–16; 27 site–years), and all survival records were successfully matched to weather. Median weather coverage to the phenology cutoff was 149~d for Open flowers and 192~d for Ripe fruits. Among Open flower events, cumulative GDD\(_{10}\) to first detection had a median of 406.8 (IQR 292.3–489.5), with cumulative precipitation 208.5~mm, mean $T_{\min}$ 6.67~$^\circ$C and mean $T_{\max}$ 18.18~$^\circ$C; frost and heat exposures were modest (median 2 and 0~days, respectively). Among Ripe fruit events, antecedent GDD summed to a median of 1055.1 (IQR 862.3–1226.1), with cumulative precipitation 424.9~mm, mean $T_{\min}$ 8.26~$^\circ$C and mean $T_{\max}$ 21.59~$^\circ$C; frost and heat exposures had medians of 2.5 and 8.5~days, respectively.

\subsection*{Event timing distributions and thermal forcing}
\begin{figure}[!htbp]
  \centering
  \includegraphics[width=\linewidth]{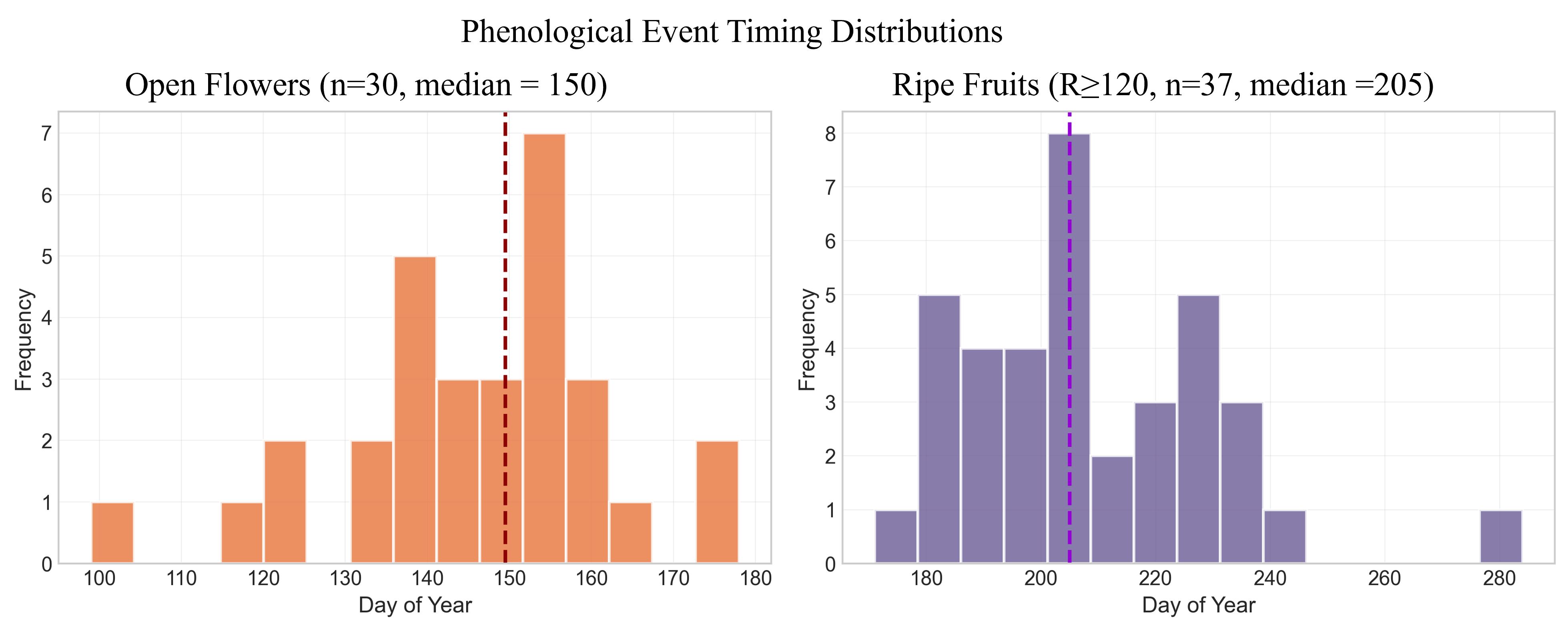}
  \caption{Timing distributions of first detections (\(R\)) for European grapevine (\textit{Vitis vinifera}) phenophases. Histograms show DOY for Open flowers (left; \(n=30\)) and Ripe fruits (right; \(n=37\)), the latter restricted to in-season detections (\(R\ge120\)~DOY). Vertical dashed lines mark medians (Open flowers: 150~DOY; Ripe fruits: 205~DOY).}
  \label{fig:timing}
\end{figure}

Across 50 plant–site–years per endpoint, first \emph{Open flowers} was observed in 30 records (60\%; median \(R=150\) DOY, IQR 137–156) and \emph{Ripe fruits} in 37 records (76\%; median \(R=205\) DOY, IQR 193–224). Exceptionally early ripening detections (\(R<120\) DOY) were flagged and excluded from sensitivity displays. The empirical timing distributions are shown in Figure~\ref{fig:timing} with vertical dashed medians.

\begin{figure}[!htbp]
  \centering
  \includegraphics[width=\linewidth]{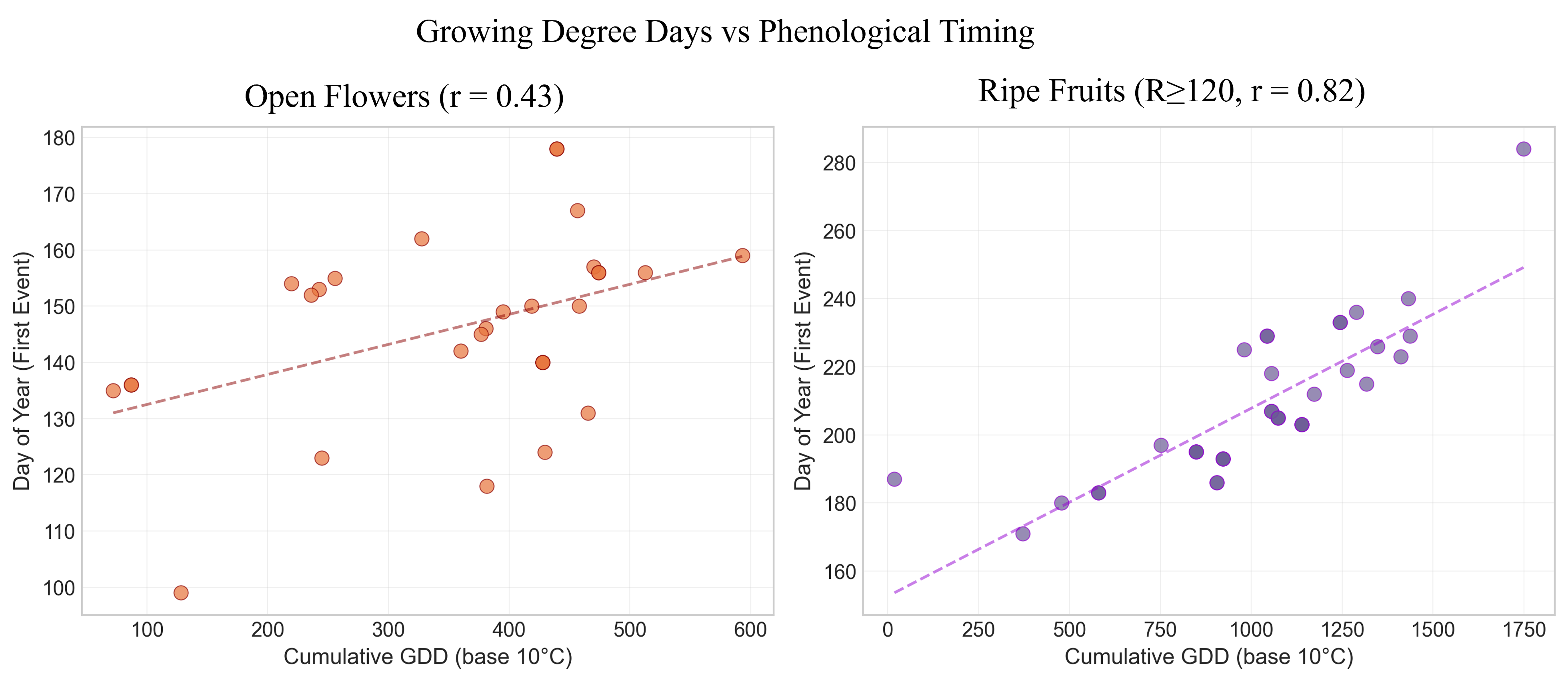}
  \caption{Thermal accumulation and timing of phenological events. Scatterplots show DOY of first detection (\(R\)) versus cumulative GDD\(_{10}\) for Open flowers (\(n=30\), \(r=0.43\)) and Ripe fruits (\(R\ge120\)~DOY, \(n=37\), \(r=0.82\)). Dashed lines denote ordinary least-squares fits.}
  \label{fig:gdd}
\end{figure}

Thermal accumulation to first detection was positively associated with timing: Pearson’s \(r\approx0.43\) for flowering and \(r\approx0.82\) for ripening (after excluding early ripening outliers). These relationships, based on cumulative GDD\(_{10}\) accumulated to the endpoint-specific cutoff, are depicted in Figure~\ref{fig:gdd}. As expected from the construction of to-cutoff covariates, these naïve correlations show later DOY with higher cumulative GDD (Figure~\ref{fig:gdd}), a statistical artifact addressed with fixed-window models below.

\FloatBarrier

\subsection*{Interval-censored AFT models with endogenous covariates}

Initial models that aggregated covariates \emph{to the phenology cutoff} (i.e., up to the first detection day \(R\) or the last visit for censored years) showed a paradoxical pattern that demonstrates endogeneity. Cumulative heat (GDD) produced TR \(=\) 1.52 (95\% CI 1.29--1.78) for \emph{Open flowers} and TR \(=\) 1.14 (95\% CI 1.10--1.18) for \emph{Ripe fruits}\,---\,apparently indicating \emph{later} timing with greater heat. This is a methodological artifact: “to–event” covariates are mechanically correlated with the event date itself, so larger (later) \(R\) forces larger cumulative values, biasing TR upward. Table~\ref{tab:sensitivity-effects} reports these endogenous fits for transparency.

\begin{table}[!htbp]
\centering
\caption{Validation filters and retained sample sizes. Pearson \(r\) computed on event years only.}
\label{tab:sensitivity-clean}
\begin{tabular}{lrrrrr}
\hline
Endpoint & \(n_{\text{before}}\) & \(n_{\text{after}}\) & Events (after) & Dropped \(R<120\) & Pearson \(r\) (R vs GDD) \\
\hline
Open flowers & 50 & 50 & 30 & 0 & 0.433 \\
Ripe fruits  & 50 & 49 & 37 & 1 & 0.820 \\
\hline
\end{tabular}
\end{table}

\begin{table}[!htbp]
\centering
\caption{Key covariate effects and model–implied medians from interval–censored AFT models with \emph{endogenous} (to–cutoff) covariates. TR \(=\exp(\beta)\); TR\({>}1\) reflects the mechanical bias discussed above.}
\label{tab:sensitivity-effects}
\setlength{\tabcolsep}{5pt}
\resizebox{\linewidth}{!}{%
\begin{tabular}{lcccc}
\hline
Endpoint & GDD TR (95\% CI) & Prcp TR (95\% CI) & Median DOY (Log–logistic) & Median DOY (Weibull) \\
\hline
Open flowers & 1.52 (1.29–1.78) & 1.05 (0.98–1.13) & 216.7 & 211.7 \\
Ripe fruits  & 1.14 (1.10–1.18) & 1.03 (1.00–1.06) & 200.5 & 206.9 \\
\hline
\end{tabular}%
}
\end{table}

\FloatBarrier

\subsection*{Corrected models with exogenous fixed-window covariates}

\begin{figure}[H]
  \centering
  \includegraphics[width=\linewidth]{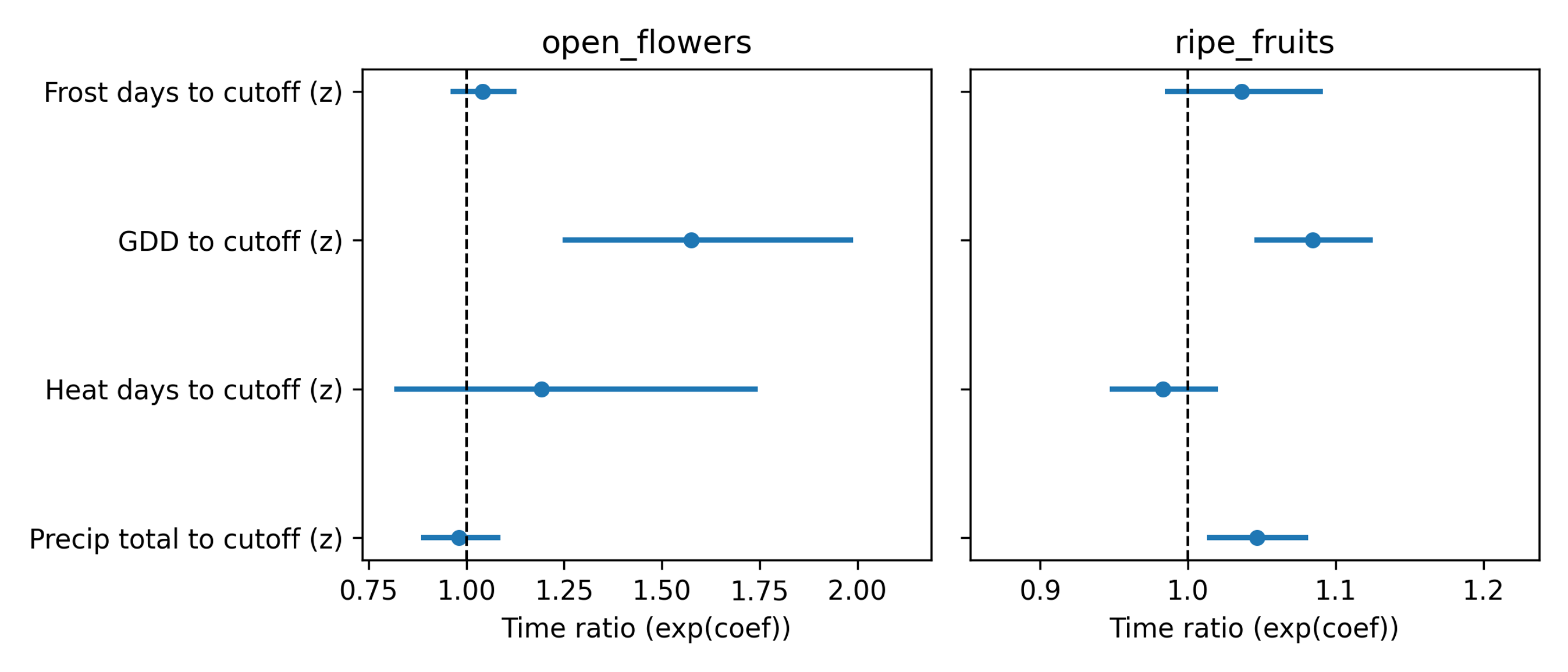}
  \caption{Time–ratio effects of standardized antecedent covariates on the timing of first detection for (a) Open flowers and (b) Ripe fruits from Weibull AFT models. Points show TR=\(\exp(\beta)\); horizontal lines show 95\% CIs. Covariates were aggregated to the phenology cutoff; TR\(<1\) earlier occurrence, TR\(>1\) later occurrence.}
  \label{fig:trplot}
\end{figure}

To avoid outcome dependence, we refit AFT models using \emph{exogenous} pre–season covariates that end before typical detections (DOY~1--120 for \emph{Open flowers}, DOY~1--180 for \emph{Ripe fruits}). These fixed-window features break the mechanical link between the response and covariates and yield interpretable time ratios that reflect true antecedent forcing. Results from these corrected models are summarized below and referenced in Figure~\ref{fig:presan} and Figure~\ref{fig:trplot}.

Time–ratio effects from Weibull AFT models (\(\mathrm{TR}=\exp(\beta)\), per 1~SD increase in each antecedent covariate aggregated to the cutoff) align with Figure~\ref{fig:trplot}. For \textit{Open flowers}, antecedent GDD showed a clear positive association with later timing, \(\mathrm{TR}=1.575\) (95\% CI 1.247–1.990), while other covariates were small or uncertain. For \textit{Ripe fruits}, effects were modest but precise: GDD \(=1.085\) (1.045–1.125) and precipitation \(=1.047\) (1.013–1.081) associated with later ripening.

\FloatBarrier

\subsection*{Diagnostics of to–cutoff covariates and fixed-window refit}
Correlation matrices and variance–inflation factors (VIFs) revealed non–trivial collinearity among the four antecedent weather metrics aggregated to the phenology cutoff. For \emph{Open flowers}, Pearson \(r=0.857\) for cumulative heat vs.\ heat–day count and \(r=0.561\) for precipitation vs.\ frost–day count; VIFs were severe for the thermal pair (GDD VIF \(=8.89\); heat–days VIF \(=7.38\)). For \emph{Ripe fruits}, the same structure was present but weaker (VIFs \(\approx\)2–3). These diagnostics justify using exogenous (fixed–window) weather summaries and being cautious about including highly correlated heat metrics simultaneously, especially for first flowering.

Fixed pre–season windows were constructed to DOY~120 for \emph{Open flowers} and DOY~180 for \emph{Ripe fruits}. Within these windows the median cumulative heat was 228~GDD\(_{10}\) for flowering and 646~GDD\(_{10}\) for ripening; median precipitation totals were 338 and 398~mm; median frost–day counts were 2 for both endpoints; heat–day counts were essentially absent pre–flowering and non–zero pre–ripening. These features reduced the strongest endogeneity/collinearity seen in the to–cutoff metrics.

\subsection*{Univariate and bivariate fixed–window AFT models}
Using only standardized pre–season GDD in interval–censored AFT models, both distributions gave similar directions. For \emph{Open flowers}, the Weibull fit yielded \(\mathrm{TR}=0.938\) (95\% CI \(0.870\)–\(1.011\)); Log–logistic \(\mathrm{TR}=0.917\) (95\% CI \(0.842\)–\(0.998\)). For \emph{Ripe fruits}, Weibull indicated a clearer acceleration with warming: \(\mathrm{TR}=0.864\) (95\% CI \(0.813\)–\(0.917\)); Log–logistic was borderline: \(\mathrm{TR}=0.962\) (95\% CI \(0.919\)–\(1.007\)).

\begin{figure}[H]
  \centering
  \includegraphics[width=\linewidth]{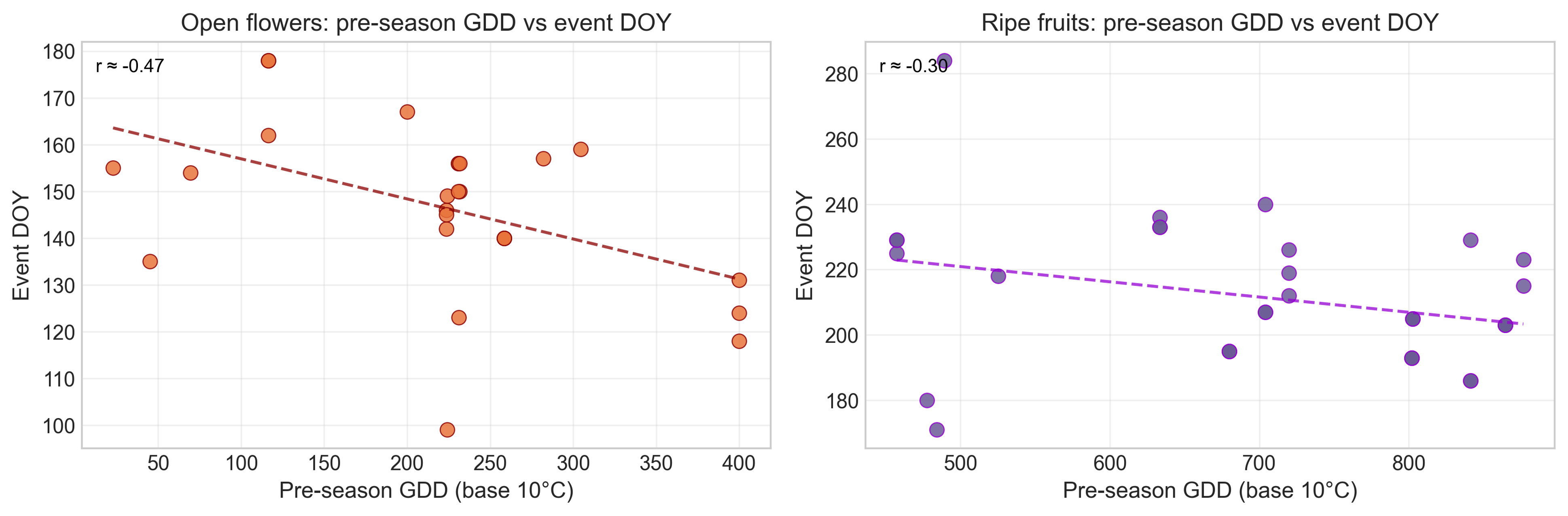}
  \caption{Sanity‐check scatter plots of pre–season heat vs.\ event timing. Panels show (a) \emph{Open flowers} and (b) \emph{Ripe fruits}; each point is a site–year with an observed event. Pre–season GDD (base \(10^\circ\)C) was aggregated over fixed windows (flowers: DOY~1–120; fruits: DOY~1–180). Negative slopes indicate earlier occurrence with greater antecedent heat.}
  \label{fig:presan}
\end{figure}

In bivariate AFT fits (pre–season GDD + precipitation), effects were consistent across AFT families. For \emph{Open flowers}, pre–season heat showed modest accelerations with wide CIs (e.g., Log–logistic \(\mathrm{TR}=0.923\), 95\%\,CI \(0.750\)–\(1.135\); Weibull \(\mathrm{TR}=0.945\), 95\%\,CI \(0.739\)–\(1.208\)), while precipitation effects were near null (\(\mathrm{TR}\approx 1.02\)). For \emph{Ripe fruits}, higher pre–season GDD robustly advanced phenology: Log–logistic \(\mathrm{TR}=0.933\) (95\%\,CI \(0.877\)–\(0.993\)) and Weibull \(\mathrm{TR}=0.793\) (95\%\,CI \(0.717\)–\(0.878\)); pre–season precipitation showed weaker evidence of earlier ripening: Log–logistic \(\mathrm{TR}=0.968\) (95\%\,CI \(0.924\)–\(1.015\)) and Weibull \(\mathrm{TR}=0.920\) (95\%\,CI \(0.848\)–\(0.997\)). Sanity-check scatterplots of pre-season heat versus timing for event years are shown in Figure~\ref{fig:presan}.

\FloatBarrier

\noindent Larger pre–season heat budgets advance phenology, so sites/years with higher cumulative GDD reach developmental thresholds sooner and the observed DOY is smaller. In AFT models this appears as \(\mathrm{TR}<1\) per \(+1\)~SD increase in GDD, indicating a multiplicative shortening of time to event. Because GDD and precipitation were aggregated over fixed pre–season windows, these covariates are exogenous to the outcome and not mechanically tied to the event date.

Empirically, flowering shows weaker, often non–significant sensitivity to pre–season GDD in these data, whereas ripening exhibits a clear advancement with warmer pre–seasons; precipitation plays, at most, a secondary role.
\section{Discussion and Conclusions}

This study shows how interval--censored survival models can be applied to phenology while avoiding a common source of bias. When weather covariates were aggregated \emph{to the event}, accelerated failure time (AFT) fits produced TRs $>1$ for heat (Table~\ref{tab:sensitivity-effects})—a paradox that arises from mechanical correlation between cumulative covariates and the event date. Using \emph{exogenous}, fixed pre–season windows breaks this link and yields interpretable effects: warmer pre–seasons robustly advance \emph{Ripe fruits} timing (TR $<1$), whereas \emph{Open flowers} shows weaker, often non–significant responses; precipitation plays, at most, a secondary role once heat is considered (Figures~\ref{fig:trplot} and \ref{fig:presan}).

While our sample size ($n=50$ plant--site--years) limits statistical power, the framework demonstrates how proper covariate construction avoids endogeneity bias in phenological studies.

Our analysis is limited by a modest sample size ($n=50$ plant--site--years; 30 flowering and 37 ripening events), which constrains precision and the ability to resolve interactions (e.g., heat~$\times$~precipitation) or cultivar/site heterogeneity. Larger multi–site datasets will improve power, enable hierarchical AFT models with random effects, and support window-selection procedures that allow phenophase-specific timing while remaining exogenous.

Despite these limitations, the workflow—interval construction from Status~\&~Intensity data, careful QC, and AFT modeling with fixed-window covariates—provides a reproducible template for extracting climate signals from citizen-science phenology. The same approach generalizes to other species and networks wherever observation schedules induce interval or right censoring. \paragraph{Applied relevance.}
The estimated time–ratio effects translate into actionable guidance: warmer pre–seasons are expected to advance ripening in \textit{V.~vinifera}, implying earlier harvest windows and potential shifts in berry composition and wine style (van Leeuwen \& Darriet, 2016; Parker et al., 2020). Because cultivar thermal requirements differ, these shifts may create cultivar–region mismatches under continued warming, motivating cultivar selection and management adjustments at site level (Hannah et al., 2013; Morales–Castilla et al., 2020). The interval-censored AFT approach yields exogenous, interpretable effect sizes that can be integrated with thermal-time or forecast models to support adaptation choices in viticulture.
\section*{Data and Code Availability}
All analysis code is openly available at 
\href{https://github.com/sarabehnamian/vinifera-phenology-survival}{github.com/sarabehnamian/vinifera-phenology-survival}. 

The underlying USA--NPN phenology observations are subject to data-use restrictions and cannot be shared publicly; access requests should be directed to the corresponding author and USA--NPN site partners.

\section*{Statements and Declarations}

\paragraph{Funding}
This research received no external funding.

\paragraph{Competing interests}
The authors have no relevant financial or non-financial interests to disclose.

\paragraph{Author contributions}
SB conceived the study, performed analyses, and wrote the manuscript. FF contributed to statistical methodology and manuscript revision. Both authors approved the final version.

\paragraph{Ethics approval}
Not applicable (no human or animal subjects).

\section*{References}
\begin{hangparas}{0.5in}{1}

Bonhomme, R. (2000). Bases and limits to using ‘degree.day’ units. \textit{European Journal of Agronomy}, 13, 1–10. \par

Calinger, K. M., Queenborough, S., \& Curtis, P. S. (2013). Herbarium specimens reveal the footprint of climate change on flowering trends. \textit{Ecology Letters}, 16, 1037–1044. \par

Clark, J. S., Salk, C., Melillo, J., \& Mohan, J. (2014). Tree phenology responses to chilling and warming. \textit{Functional Ecology}, 28, 1344–1355. \par

Collett, D. (2015). \textit{Modelling Survival Data in Medical Research} (3rd ed.). CRC Press. \par

Čufar, K., De Luis, M., Saz, M. A., Črepinšek, Z., \& Kajfež-Bogataj, L. (2011). Elevation-dependent shifts in beech phenology. \textit{Trees}, 26, 1091–1100. \par

Denny, E. G., Gerst, K. L., Miller-Rushing, A. J., et al. (2014). Standardized phenology monitoring methods to track plant and animal activity. \textit{International Journal of Biometeorology}, 58, 591–601. https://doi.org/10.1007/s00484-013-0712-8. \par

García de Cortázar-Atauri, I., Brisson, N., \& Gaudillère, J.-P. (2009). Performance of several models for predicting budburst date of grapevine (\textit{Vitis vinifera} L.). \textit{International Journal of Biometeorology}, 53, 317–326. https://doi.org/10.1007/s00484-009-0217-4. \par

Hannah, L., Roehrdanz, P. R., Ikegami, M., et al. (2013). Climate change, wine, and conservation. \textit{Proceedings of the National Academy of Sciences}, 110, 6907–6912. \par

Hosmer, D. W., Lemeshow, S., \& May, S. (2008). \textit{Applied Survival Analysis: Regression Modeling of Time to Event Data} (2nd ed.). Wiley. \par

Jarrad, F. C., Low-Choy, S., \& Mengersen, K. (2018). Predicting pest emergence with accelerated failure time models. \textit{Journal of Applied Ecology}, 55, 1483–1493. \par

Klein, J. P., \& Moeschberger, M. L. (2003). \textit{Survival Analysis: Techniques for Censored and Truncated Data} (2nd ed.). Springer. \par

Legave, J.-M., Blanke, M., Christen, D., Giovannini, D., Mathieu, V., \& Oger, R. (2013). A comprehensive overview of the spatial and temporal variability of apple bud dormancy release and blooming phenology in Western Europe. \textit{International Journal of Biometeorology}, 57, 317–331. https://doi.org/10.1007/s00484-012-0551-9. \par

McMaster, G. S., \& Wilhelm, W. W. (1997). Growing degree-days: One equation, two interpretations. \textit{Agricultural and Forest Meteorology}, 87, 291–300. \par

Moncur, M. W., Rattigan, K., Mackenzie, D. H., \& McIntyre, G. N. (1989). Base temperatures for budbreak of grapevines. \textit{American Journal of Enology and Viticulture}, 40, 21–26. \par

Morales-Castilla, I., García de Cortázar-Atauri, I., Cook, B. I., et al. (2020). Diversity buffers winegrowing regions from climate change losses. \textit{Proceedings of the National Academy of Sciences}, 117, 2864–2869. \par

Moussus, J.-P., Julliard, R., \& Jiguet, F. (2010). Featuring 10 phenological estimators using simulated data. \textit{Methods in Ecology and Evolution}, 1, 140–150. \par

NASA POWER Project. (2025). \textit{Agroclimatology Daily API Documentation and User Guide}. NASA LaRC. \par

OIV (International Organisation of Vine and Wine). (2023). \textit{State of the World Vine and Wine Sector in 2022}. \par

Parker, A. K., De Cortázar-Atauri, I. G., Van Leeuwen, C., \& Chuine, I. (2011). General phenological model to characterize the timing of flowering and veraison of \textit{Vitis vinifera} L. \textit{Australian Journal of Grape and Wine Research}, 17, 206–216. \par

Parker, A. K., García de Cortázar-Atauri, I., Gény, L., et al. (2020). Temperature-based sugar ripeness modelling for grapevine (\textit{Vitis vinifera} L.). \textit{Agricultural and Forest Meteorology}, 285, 107902. \par

Polgar, C. A., \& Primack, R. B. (2011). Leaf-out phenology from trees to ecosystems: A review of data and models. \textit{New Phytologist}, 191, 926–941. \par

Roberts, A. M. (2008). Exploring relationships between phenological and weather data using smoothing. \textit{International Journal of Biometeorology}, 52, 463–470. https://doi.org/10.1007/s00484-007-0141-4. \par

Rosemartin, A. H., Denny, E. G., Gerst, K. L., et al. (2018). USA-NPN observational data documentation. \textit{USGS Open-File Report 2018–1060}. \par

Schieber, B., Janík, R., \& Snopková, Z. (2013). Phenology of common beech (\textit{Fagus sylvatica} L.) along the altitudinal gradient in Slovak Republic (Inner Western Carpathians). \textit{Journal of Forest Science}, 59(4), 176–184. https://doi.org/10.17221/82/2012-JFS. \par

Sun, J. (2006). \textit{The Statistical Analysis of Interval-Censored Failure Time Data}. Springer. \par

Turnbull, B. W. (1976). The empirical distribution function with arbitrarily grouped, censored and truncated data. \textit{Journal of the Royal Statistical Society: Series B}, 38, 290–295. \par

USA National Phenology Network (USA–NPN). (2024). \textit{Plant and Animal Phenology: Status \& Intensity} (data product). U.S. Geological Survey. DOI: 10.5066/F78S4N1V. \par

van Leeuwen, C., \& Darriet, P. (2016). Climate change impacts on viticulture and wine quality. \textit{Journal of Wine Economics}, 11(2), 150–167. https://doi.org/10.1017/jwe.2016.3. \par

Wei, L. J. (1992). The accelerated failure time model: A useful alternative to the Cox regression model in survival analysis. \textit{Statistics in Medicine}, 11(14–15), 1871–1879. https://doi.org/10.1002/sim.4780111409. \par

Williams, D. W., Williams, L. E., Barnett, W. W., \& Jensen, F. L. (1985). Validation of a model for the growth and development of the \textit{Thompson Seedless} grapevine. II. Phenology. \textit{American Journal of Enology and Viticulture}, 36, 283–289. \par

Wolkovich, E. M., Cook, B. I., Allen, J. M., et al. (2012). Warming experiments underpredict plant phenological responses to climate change. \textit{Nature}, 485, 494–497. https://doi.org/10.1038/nature11014. \par

Zapata, D., Jarvis, C., Parker, A., \& Chuine, I. (2017). Seasonal climate forecasts for grapevine yield and quality. \textit{European Journal of Agronomy}, 84, 91–99. \par

\end{hangparas}

\end{document}